# X-ray harmonic comb from relativistic electron spikes

Alexander S. Pirozhkov[1], Masaki Kando[1], Timur Zh. Esirkepov[1], Eugene N. Ragozin[2,3], Anatoly Ya. Faenov[1,4], Tatiana A. Pikuz[1,4], Tetsuya Kawachi[1], Akito Sagisaka[1], Michiaki Mori[1], Keigo Kawase[1], James K. Koga[1], Takashi Kameshima[1], Yuji Fukuda[1], Liming Chen[1,†], Izuru Daito[1], Koichi Ogura[1], Yukio Hayashi[1], Hideyuki Kotaki[1], Hiromitsu Kiriyama[1], Hajime Okada[1], Nobuyuki Nishimori[1], Kiminori Kondo[1], Toyoaki Kimura[1], Toshiki Tajima[1,§], Hiroyuki Daido[1], Yoshiaki Kato[1,‡] & Sergei V. Bulanov[1,5]

*[1]Advanced Photon Research Center, Japan Atomic Energy Agency, 8-1-7 Umemidai, Kizugawa-shi, Kyoto 619-0215, Japan; [2]P. N. Lebedev Physical Institute of the Russian Academy of Sciences, Leninsky Prospekt 53, 119991 Moscow, Russia; [3]Moscow Institute of Physics and Technology (State University), Institutskii pereulok 9, 141700 Dolgoprudnyi, Moscow Region, Russia; [4]Joint Institute of High Temperatures of the Russian Academy of Sciences, Izhorskaja Street 13/19, 127412 Moscow, Russia; [5]A. M. Prokhorov Institute of General Physics of the Russian Academy of Sciences, Vavilov Street 38, 119991 Moscow, Russia*

[†]Present address: Institute of Physics of the Chinese Academy of Sciences, Beijing, China. [§]Present address: Ludwig-Maximilians-University, Germany. [‡]Present address: The Graduate School for the Creation of New Photonics Industries, 1955-1 Kurematsu-cho, Nishiku, Hamamatsu, Shizuoka, 431-1202, Japan.

**X-ray devices providing nanometre spatial[1] and attosecond[2] temporal resolution are far superior to longer wavelength and lower frequency optical ones. Such resolution is indispensable in biology, medicine, physics, material sciences, and their applications. A bright ultrafast coherent X-ray source is highly desirable, as**

its single shot[3] would allow achieving high spatial and temporal resolution simultaneously, which are necessary for diffractive imaging of individual large molecules, viruses, or cells. Here we demonstrate experimentally a new compact X-ray source involving high-order harmonics produced by a relativistic-irradiance femtosecond laser in a gas target. In our first implementation using a 9 Terawatt laser, coherent soft X-rays are emitted with a comb-like spectrum reaching the 'water window' range. The generation mechanism is robust being based on phenomena inherent in relativistic laser plasmas: self-focusing,[4] nonlinear wave generation accompanied by electron density singularities,[5] and collective radiation by a compact electric charge proportional to the charge squared. The formation of singularities (electron density spikes) is described by the elegant mathematical catastrophe theory,[6] which explains sudden changes in various complex systems, from physics to social sciences. The new X-ray source has advantageous scalings, as the maximum harmonic order is proportional to the *cube* of the laser amplitude enhanced by relativistic self-focusing in plasma. This allows straightforward extension of the coherent X-ray generation to the keV and tens of keV spectral regions with 100 Terawatt and Petawatt lasers, respectively. The implemented X-ray source is remarkably easily accessible: the requirements for the laser can be met in a university-scale laboratory, the gas jet is a replenishable debris-free target, and the harmonics emanate directly from the gas jet without additional devices. Our results open the way to a compact coherent ultrashort brilliant X-ray source with single shot and high-repetition rate capabilities, suitable for numerous applications and diagnostics in many research fields.

High-order harmonic generation is the manifestation of one of the most fundamental properties of nonlinear wave physics. Numerous examples of nonlinearities producing high-frequency waves can be seen in everyday life: a whistle, where a continuous air flow is converted into high-frequency sound, a human voice and musical instruments, where harmonics called overtones enrich sound imparting uniqueness and beauty. In intense laser-matter interactions, nonlinear electromagnetic waves produce harmonics which are used as coherent radiation sources in many applications; studying harmonics inspires new concepts of nonlinear wave theory.

Several compact laser-based X-ray sources have been implemented to date, including plasma-based X-ray lasers,[7] atomic harmonics in gases,[8] nonlinear Thomson scattering[9, 10] from plasma electrons[11, 12] and electron beams,[13, 14] harmonics from solid targets,[15-21] and relativistic flying mirrors.[22-25] Many of these compact X-ray sources[9-25] are based on the relativistic laser-matter interaction, where the dimensionless laser pulse amplitude $a_0 = eE_0/mc\omega_0 = (I_0/I_{rel})^{1/2}(\lambda_0/\mu m) > 1$. Here $e$ and $m$ are the electron charge and mass, $c$ is the speed of light in vacuum, $\omega_0$, $\lambda_0$, $E_0$, and $I_0$ are the laser angular frequency, wavelength, peak electric field, and peak irradiance, respectively, and $I_{rel} = 1.37 \times 10^{18}$ W/cm$^2$. Such a relativistic-irradiance laser pulse ionizes matter almost instantly, so the interaction takes place in plasma, which can sustain extremely high laser irradiance. This, in particular, allows generating very high harmonic orders in an ultra-relativistic regime ($a_0 \gg 1$). Recently a great deal of attention has been paid to harmonic generation from solid targets.[17-21] Gas targets, which are easily accessible and far less demanding with respect to the laser contrast, also have been employed to generate harmonics via electro-optic shocks[26] and nonlinear Thomson scattering.[11, 12] In our experiments, a relativistic-irradiance laser focused into a gas jet generates bright harmonics having a large number of photons with energies of hundreds of electron-Volts (eV). A new mechanism is invoked for explaining the obtained results.

A laser pulse with the power $P_0 = 9$ TW, duration of 27 fs, and wavelength $\lambda_0 = 820$ nm is focused onto a supersonic helium gas jet. The laser irradiance in vacuum is $6.5 \times 10^{18}$ W/cm$^2$, corresponding to the dimensionless amplitude $a_{0,vac} \approx 1.7$. The harmonics in the 80-250 eV or 110-350 eV spectral regions have been recorded in the forward (laser propagation) direction employing a flat-field grazing-incidence spectrograph.

The harmonics are generated in a broad range of plasma densities from ~ $2 \times 10^{19}$ to $7 \times 10^{19}$ cm$^{-3}$, with the harmonic yield increasing with density. Greater harmonic yield is achieved at higher laser power by shortening the pulse duration while maintaining the laser pulse energy. This is advantageous as it allows using more compact laser systems.

Figure 1 shows a typical comb-like spectrum consisting of odd and even harmonics, which both are similar in intensity and shape. The base frequency of the spectrum, $\omega_f$, is downshifted compared to the laser frequency $\omega_0$ due to the well-known gradual downshift of the laser pulse frequency, as the pulse propagates in tenuous plasma expending part of its energy. This frequency downshift has been observed in the present experiment by recording the transmitted laser spectra.

The large photon number allows recording spectra in a single shot. In the data shown in Fig. 1, the photon number and X-ray pulse energy within the spectral range of 90-250 eV reach $(1.8 \pm 0.1) \times 10^{11}$ photons/sr and $(3.2 \pm 0.2)$ μJ/sr, respectively. For the harmonic source peak brightness a conservative estimate gives ~$10^{21}$ photons/(mm$^2$ mrad$^2$ s 0.1% bandwidth) at 100 eV and ~$10^{20}$ photons/(mm$^2$ mrad$^2$ s 0.1% bandwidth) at ~200 eV, respectively; these numbers obtained in our first implementation compare well with other sources.[27]

In Fig. 1**d**, harmonic orders up to $n_H{}^* = \omega^*/\omega_f \sim 126$ are resolved, and the unresolved (continuum) spectrum extends up to 200 eV, being close to the cut-off of the optical blocking filters used in the spectrograph. With another filter set, we have observed spectra with photon energies exceeding 320 eV, well within the 'water window' range (284 – 532 eV), which is an important region for high-contrast imaging of biological samples. The large number of resolved harmonics places a strict limit on the laser frequency drift during the harmonic emission, because resolving of the $n_H$-th harmonic allows a relative change of the laser frequency not greater than $\delta \approx 1/(2n_H)$. For the case of Fig. 1, the upper bound is $\delta_{max} \sim 0.4\%$ (or, alternatively, the phase error is <25 mrad). The gradual downshift of the laser frequency in plasma thus limits the harmonic emission time. For the shot shown in Fig. 1 we obtain a conservative estimate for the emission time of $\approx 13$ fs.

In ~40% of the shots, the harmonic spectrum exhibits deep equidistant modulations (Fig. 2**a-c**), suggesting interference between two almost identical coherent sources. These modulations remain visible even in shots where the individual harmonics are nearly unresolved (Fig. 2**d**). We attribute harmonic structure blurring to a larger laser frequency drift due to longer harmonic emission time. The photon numbers in these shots are correspondingly a few times larger.

The conservative estimate of the spatial coherence width in our experiment, with the detector at 1.4 m, gives ~ 1 mm, which is large enough for phase contrast imaging in a compact setup.

The unique properties of the observed harmonics prevent direct application of previously suggested scenarios. Atomic harmonics are excluded because in the present experiment odd and even harmonic orders are always generated and the sensitivity to the gas pressure is weak. Betatron radiation consists of harmonics with a base frequency determined by the plasma frequency and electron energy, and not the laser frequency. Nonlinear Thomson scattering even under optimum conditions[10] can only provide photon numbers two orders of magnitude smaller than that experimentally observed.

Properties of the observed spectra and our 3-dimensional Particle-in-Cell simulation of the harmonic generation during laser pulse propagation in tenuous plasma (Fig. 3**a**) allow inferring the mechanism of harmonic generation. The laser pulse undergoes self-focusing,[4] expels electrons evacuating a cavity[5] and generates a bow wave.[28] The resulting electron flow is two-stream: expelled electrons produce the outgoing bow wave while peripheral electrons close the cavity. Fig. 3**b** illustrates this flow by the evolution of an initially flat surface formed by electrons in their phase sub-space $(x, y, p_y)$, where $p_y$ is the electron momentum component. The laser pulse stretches the surface making folds – outer (the bow wave boundary) and inner (the cavity wall). A projection of the surface onto the $(x,y)$ plane gives the electron density distribution, where a fold corresponds to the density singularity. Catastrophe theory[6] predicts here universal structurally stable singularities. The 'fold' type ($A_2$, according to V. I. Arnold's ADE classification) is observed at the bow wave boundary and at the cavity wall where the density grows as $(\Delta r)^{-1/2}$ while the distance to the singularity, $\Delta r$, diminishes. At the point joining two folds, the density grows as $(\Delta r)^{-2/3}$ producing a higher order singularity, the 'cusp' ($A_3$). Our simulation reveals the formation of the electron density spike corresponding to the cusp singularity, located in a ring surrounding the cavity head and modulated by the laser field. The density spike moving together with the laser pulse carries a large localized electric charge, collective motion of which under the action of the laser field generates high-order harmonics, Fig. 3**a**. A large concentration of electrons ensured by the cusp singularity makes the emission coherent, so that the emitted intensity is proportional to the square of the particle number, $N_e^2$. The estimated number of electrons within the singularity ring, $N_e \sim 10^6$, provides the signal level similar to the experiment. For linear polarization, the harmonic emitting ring breaks up into two segments (Fig. 3). In the symmetric case, these segments radiate identical spectra, interference between which explains modulations visible in Fig. 2. If the symmetry is violated, as in Fig. 3**a**, the radiation from one source dominates, as in the shot shown in Fig. 1.

The critical harmonic order, $n_H^*$, is proportional to the cube of the electron energy ($\approx a_0 mc^2$), similar to synchrotron radiation and nonlinear Thomson scattering:

$$n_H^* = \frac{\omega^*}{\omega_l} \sim a_0^3. \tag{1}$$

Estimating the laser amplitude $a_0$ in the stationary self-focusing case,[29] $a_0 = [8\pi P_0 n_e/(P_c n_{cr})]^{1/3}$, the critical harmonic order becomes $n_H^* \sim 8\pi P_0 n_e/(P_c n_{cr})$, where $P_c = 2m^2 c^5/e^2 = 17$ GW and $n_{cr} = m\omega_0^2/(4\pi e^2) \approx 1.1 \times 10^{21}$ cm$^{-3}$($\mu$m/$\lambda_0$)$^2$ is the critical density. The observed harmonic orders are in good agreement with this scaling. The total energy emitted by the cusp is proportional to the charge squared:

$$W \approx \frac{e^2}{8c} N_e^2 a_0^4 \gamma \omega_0^2 \tau \propto N_e^2 P_0^{4/3} n_e^{5/6} \omega_0^{1/3} \tau, \tag{2}$$

where $\gamma \approx (n_{cr}/n_e)^{1/2}$ is the gamma-factor associated with the group velocity of the laser pulse in plasma. The number of electrons $N_e$ and the harmonic emission time $\tau$ depend on the detailed structure of the cusp and its time evolution.

Using a compact femtosecond laser and relatively simple setup with a replenishable, debris-free gas jet target suitable for repetitive operation, we demonstrate a bright, coherent X-ray source with advantageous properties such as scalability in photon energy and number, single-shot capability, and femtosecond pulse duration. Our findings have immediate applications in ultrafast science and plasma physics and in the near future will impact other fields of science, medicine, and technology, where convenient X-ray sources are required, including those with time-resolved and single-shot capabilities.

We acknowledge the technical support of Drs. Takayuki Homma, Jinglong Ma, and the J-KAREN laser team. We acknowledge the financial support from MEXT (Kakenhi 20244065 and 19740252) and JAEA President Grants.

Correspondence and requests for materials should be addressed to A.S.P.

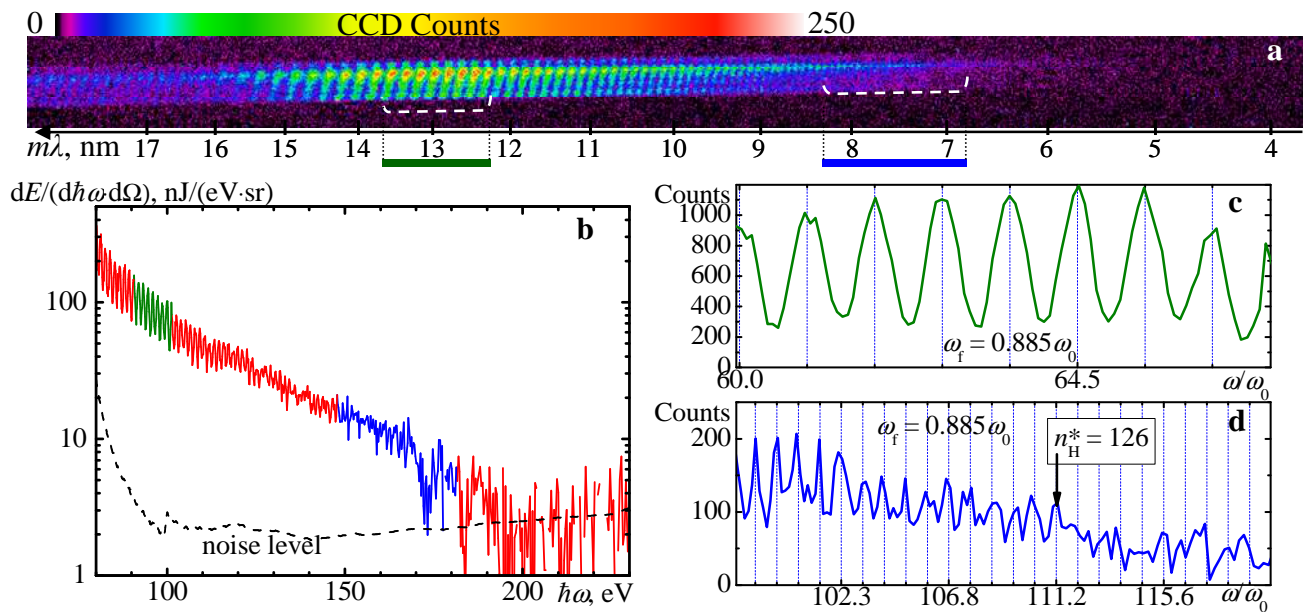

Fig. 1. A typical single-shot harmonic spectrum, the electron density is 2.7×10¹⁹ cm⁻³. **a** A portion of the raw data recorded with a CCD. **b** The spectrum obtained from the raw data line-out taking into account the toroidal mirror reflectivity, filter transmission, grating efficiency, and CCD quantum efficiency. The "noise level" (dashed curve) includes the CCD dark current, read-out noise, and shot noise. **c**, **d** The line-outs of two selected regions (denoted by the dashed brackets in frame **a** and the corresponding colours in frames **a** and **b**) demonstrating harmonic structure with the base frequency $\omega_f = 0.885\omega_0$ denoted by the dotted vertical lines, where $\omega_0$ is the laser frequency, and the highest distinctly resolved order $n^* = \omega^*/\omega_f = 126$.

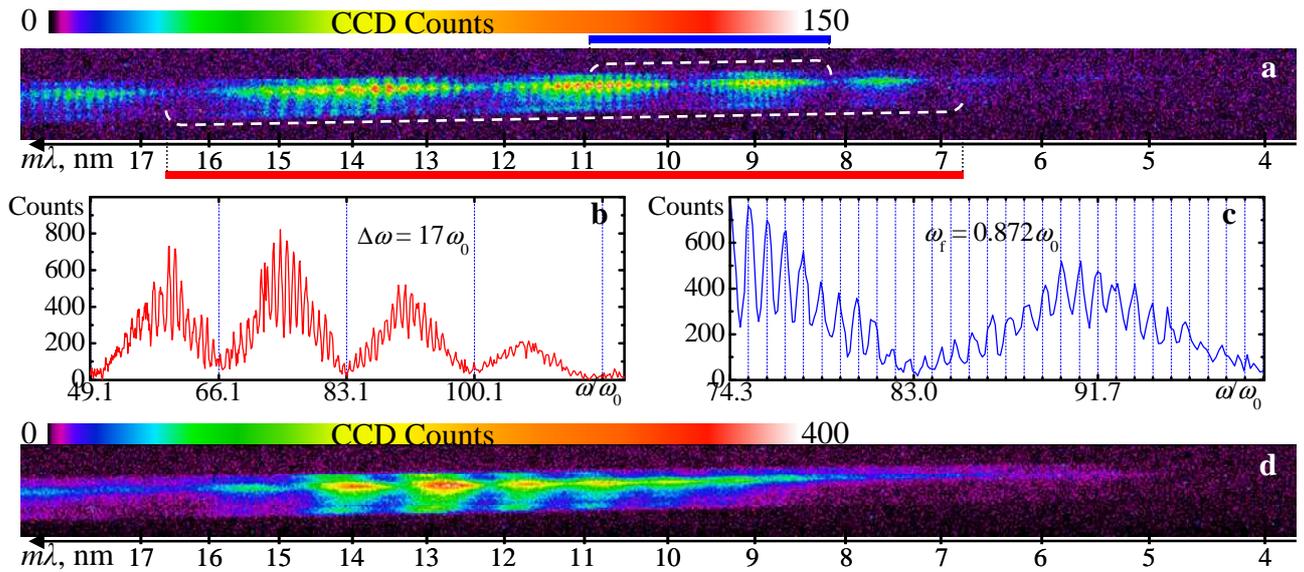

Fig. 2.  Typical single-shot spectra modulated due to the interference between two nearly identical sources (Fig. 3**b**). **a** A portion of the CCD raw data; modulated spectrum with resolved harmonics, the electron density is $4.7 \times 10^{19}$ cm$^{-3}$. **b** and **c**, line outs of two selected regions (denoted by the two dashed brackets and the corresponding colours in frame **a**). The modulation period in the frequency domain $\Delta\omega = 17\omega_0$ and the base frequency $\omega_f = 0.872\omega_0$ are denoted by the dotted vertical lines in frames **b** and **c**, respectively. **d** A portion of the CCD raw data; modulated spectrum with nearly unresolved harmonics, the electron density is $4.7 \times 10^{19}$ cm$^{-3}$.

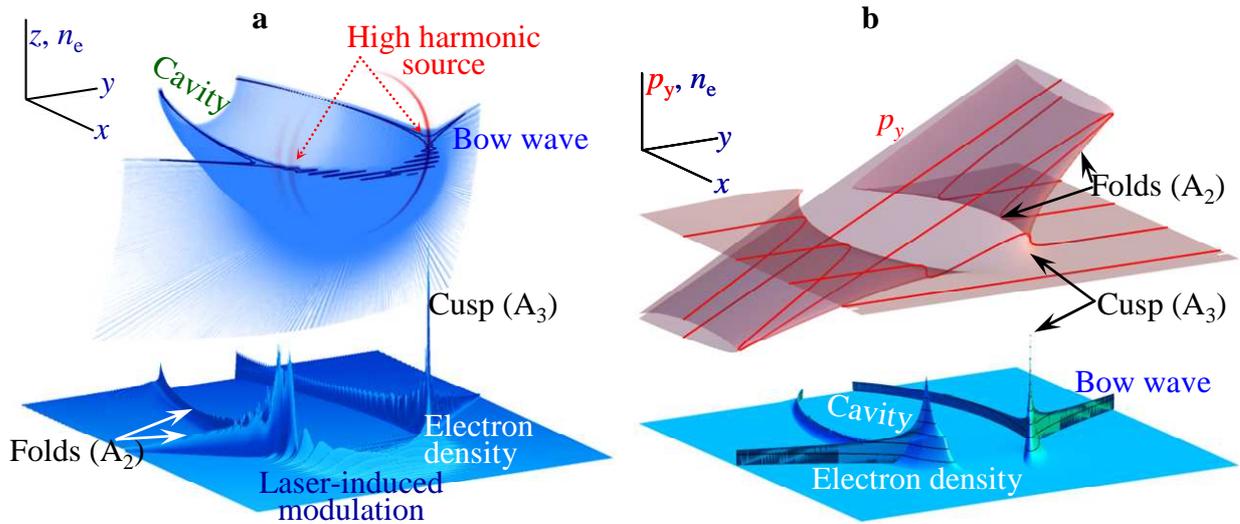

Fig. 3. Mechanism of harmonic generation, three-dimensional Particle-in-Cell simulation (**a**) and model (**b**). Electrons initially located in the plane ($x$,$y$) form a flat surface in the electron phase sub-space ($x$,$y$, $p_y$) (Fig. 3**b**, upper frame), where $p_y$ is the electron momentum component. Near the axis, the laser pulse stretches the surface making folds so that outer folds represent the bow wave[28] boundary, inner folds represent the cavity walls. A projection of the surface onto the ($x$,$y$) plane gives the electron density distribution, where according to catastrophe theory[6] the folds correspond to singularities of the density (Fig. 3 **a** and **b**). Higher order singularities, the cusps, are seen in Fig. 3 **a** and **b** at the locations of the joining of these folds. The cusps provide a large, localized electric charge. This charge is situated well within the region where the laser pulse amplitude is large, thus its nonlinear oscillations produce high-order harmonics. The spatial distribution of the electromagnetic field of the fourth and higher harmonics is shown by the red arcs in Fig. 3**a**.

# X-ray harmonic comb from relativistic electron spikes: *Supplementary Information*

Alexander S. Pirozhkov[1], Masaki Kando[1], Timur Zh. Esirkepov[1], Eugene N. Ragozin[2,3], Anatoly Ya. Faenov[1,4], Tatiana A. Pikuz[1,4], Tetsuya Kawachi[1], Akito Sagisaka[1], Michiaki Mori[1], Keigo Kawase[1], James K. Koga[1], Takashi Kameshima[1], Yuji Fukuda[1], Liming Chen[1,†], Izuru Daito[1], Koichi Ogura[1], Yukio Hayashi[1], Hideyuki Kotaki[1], Hiromitsu Kiriyama[1], Hajime Okada[1], Nobuyuki Nishimori[1], Kiminori Kondo[1], Toyoaki Kimura[1], Toshiki Tajima[1,§], Hiroyuki Daido[1], Yoshiaki Kato[1,‡] & Sergei V. Bulanov[1,5]

*[1]Advanced Photon Research Center, Japan Atomic Energy Agency, 8-1-7 Umemidai, Kizugawa-shi, Kyoto 619-0215, Japan; [2]P. N. Lebedev Physical Institute of the Russian Academy of Sciences, Leninsky Prospekt 53, 119991 Moscow, Russia; [3]Moscow Institute of Physics and Technology (State University), Institutskii pereulok 9, 141700 Dolgoprudnyi, Moscow Region, Russia; [4]Joint Institute of High Temperatures of the Russian Academy of Sciences, Izhorskaja Street 13/19, 127412 Moscow, Russia; [5]A. M. Prokhorov Institute of General Physics of the Russian Academy of Sciences, Vavilov Street 38, 119991 Moscow, Russia*

[†]Present address: Institute of Physics of the Chinese Academy of Sciences, Beijing, China. [§]Present address: Ludwig-Maximilians-University, Germany. [‡]Present address: The Graduate School for the Creation of New Photonics Industries, 1955-1 Kurematsu-cho, Nishiku, Hamamatsu, Shizuoka, 431-1202, Japan.

## Experimental Setup

The experimental setup is shown in Fig. S1**a**. The gas jet density profile and position of laser focus in vacuum are shown in Fig. S1**b**. The grazing-incidence spectrograph is in-situ calibrated using the line emission from Ar and Ne plasmas produced by the same laser in the same gas jet filled with the different gases. An example of the spectrum used for the calibration is shown in Fig. S2**a** and **b**. The resolving power of the spectrograph $\omega/\Delta\omega$ estimated from the line widths is shown in Fig. S2**c**, from which we can expect the maximum resolved harmonic order ~ 130 for the fundamental wavelength $\lambda_0 = 820$ nm and ~ 140 for the red-shifted wavelength $\lambda_0' = 927$ nm (corresponding to $\omega_f = 0.885\omega_0$, which is the base frequency of the spectrum shown in Fig. 1 of the main text). These estimations are approximate because the ability to resolve harmonics also depends on the spectral shape and the Charge-Coupled Device (CCD) noise at low signal levels. For this reason, the actually observed resolved harmonic orders up to ~ 126 may correspond to the limit imposed by the detection system.

**Image processing**

Because of the large signal-to-nose ratio, the image processing has been reduced to minimum. The raw CCD counts have been converted into pseudo-colours, with the linear colour bars shown together with each experimental data. In the raw data, there are typically several bright spots generated by hard X-rays; these bright spots have been removed by the procedure described in Ref.[30] The images before and after bright spot removal are shown in Fig. S3 **a** and **b**, respectively.

**Absolute photon number and noise calculation**

The absolute photon number in the harmonic spectra is calculated using the idealized spectrograph throughput (shown by the dashed line in Fig. S2**b**), which is the product of the toroidal mirror reflectivity (calculated using the atomic scattering factors[31, 32]), filter transmission[33] (calculated, the measured transmission at several wavelengths agrees well with the calculation), grating efficiency,[34] and CCD quantum efficiency.[35] The background has been carefully subtracted. The CCD gain $g = 0.315$ counts/electron has been measured by the manufacturer, and the energy per electron-hole pair is 3.65 eV. The effects of optic contamination by Si is partly included, calculated from the ~ 30% jump at the absorption edge near 12.4 nm observed in quasi-continuous spectra; this contamination resulted from the spectrometer operation with solid Si targets in different experiments. Other contaminations which always exist (hydrocarbon, oxygen, etc.) are not included. Neither included is absorption in the outer regions of the helium jet; this should have a small effect because the gas is ionized few picoseconds before the main pulse by a pedestal. These excluded effects (optics contamination and He absorption) can only add the brightness of the harmonics at the source, and the conclusions of the paper remain valid.

The noise (standard deviation) for each CCD pixel is calculated from the measured CCD dark and read-out noise $\sigma_d$ and the shot noise given by the product of the CCD gain and the observed counts $C$:[36]

$$\sigma^2 = \sigma_d^2 + gC \tag{S1}$$

The line outs shown in Fig. 1 and 2 of the main text are binned vertically by several pixels; the noise in each spectral point $\sigma_{bin}$ is calculated as sum of independent noise sources from each pixel: $\sigma_{bin}^2 = \sum \sigma^2$. The noise level is much lower than the typically observed signals, except at the edges of the spectrograph's throughput. The error bars for the photon numbers and energy of the harmonics in the spectral range 90-250 eV given in the main text are calculated from the photon number uncertainties at each point in the spectrum.

**Harmonics generated by the laser pulse with changing frequency**

When the driving laser frequency changes during the harmonic emission into the acceptance angle of the spectrograph, the observed time-integrated spectrum becomes blurred due to the overlapping of different harmonic orders emitted with different base frequencies at different times. A model example is shown in Fig. S4. The harmonics around order $n_H$ become distinguishable when the relative frequency change $\Delta\omega/\omega_0$ becomes smaller than

$$\frac{\Delta\omega}{\omega_0} \approx \frac{1}{2n_H} \qquad (S2)$$

For the resolvable harmonic orders up to $n_H = 126$, as in the described experiments, this gives $\Delta\omega/\omega_0 \approx 0.4\%$. Note that these estimations do not depend on the harmonic generation mechanism. In the experiments, we sometimes observed spectra without or with nearly indistinguishable harmonic structures, as it is shown in Fig. 2**d** of the main text. This is attributed to the relative frequency drift larger than $1/2n_H$.

**Laser pulse propagation in plasma: relativistic self-focusing, nonlinear depletion and frequency downshift**

A laser pulse propagating in tenuous plasma produces various nonlinear effects, which in turn influence the laser pulse itself.[37, 5] The laser pulse gradually changes due to such processes as relativistic self-focusing, wake wave excitation, etc. In particular, the relativistic self-focusing[39, 4] leads to a significant increase of the laser amplitude. In the stationary case, the amplitude and diameter of the self-focusing channel are determined by the laser power $P_0$ and electron density $n_e$[29]

$$a_0 = \left(8\pi \frac{P_0 n_e}{P_c n_{cr}}\right)^{1/3}, \; d_{sf} = 2\sqrt{a_0}\,\frac{c}{\omega_{pe}} \,. \qquad (S3)$$

Here $P_c = 2m^2c^5/e^2 \approx 17$ GW, $n_{cr} = m\omega_0^2/(4\pi e^2) \approx 1.1\times10^{21}$ cm$^{-3}$($\mu$m/$\lambda_0$)$^2$ is the critical density, $\omega_{pe} = (4\pi e^2 n_e/m)^{1/2}$ is the Langmuir frequency. For the experimental parameters of Fig. 1 of the main text ($P_0$ = 9 TW, $n_e$ = 2.7×10$^{19}$ cm$^{-3}$), we obtain $a_0$ = 6 and $d_{sf}$ = 5 µm.

Another important effect is the laser pulse frequency downshift in plasma. An example of the transmitted laser pulse spectrum is shown in Fig. S5**a**; this spectrum is obtained in the same shot as the harmonic spectrum shown in Fig. 1 of the main text. Due to this frequency downshift the base harmonic frequencies observed in the experiment are somewhat smaller than the initial laser one. The frequency downshift is attributed to the nonlinear depletion[5, 42] of the laser pulse. If the plasma density is sufficiently low, as in the case of the experiment, the energy loss rate for the wake wave excitation is relatively small so that laser pulse changes nearly adiabatically, which means that the number of photons is nearly preserved while the average frequency gradually drifts to lower values.[42] Assuming uniform plasma density, the average frequency $\omega_0'$ can be calculated as[42]

$$\frac{\omega_0'}{\omega_0} = \left(1 - \frac{x}{l_{nl}}\right)^{1/3}.$$  (S4)

Here $l_{nl} \approx \lambda_p \gamma^2$ is the nonlinearity length, $\lambda_p = 4(2\gamma)^{1/2} c/\omega_{pe}$ is the nonlinear plasma wavelength, and $\gamma$ is the gamma-factor associated with the phase velocity of the Langmuir wave, which coincides with the group velocity of the laser pulse. For the parameters of the shot shown in Fig. 1 of the main text, the electron density is $n_e$ = 2.7×10$^{19}$ cm$^{-3}$, $\lambda_p \approx$ 14 µm, and $l_{nl} \approx$ 0.5 mm. The gamma-factor $\gamma$ = 6±1 has been measured under similar experimental conditions by the frequency upshift of the reflected counter-propagating laser pulse.[25] The calculated dependence of $\omega_0'$ on the propagation distance is shown in Fig. S5**b**. As we see, the harmonics are generated by the laser pulse with gradually changing frequency. If the harmonic emission length, $\Delta x$, is sufficiently long, this frequency downshift leads to blurring of the harmonic structure, as in the case of Fig. 2**d** of the main text. Using the maximum frequency change found in the previous section and the dependence shown in Fig. S5**b**, for the shot shown in Fig. 1 of the main text we estimate the harmonic emission length as $\Delta x \approx 4$ μm, which corresponds to the harmonic emission time $\Delta x/c \approx 13$ fs. Note that this gives a very conservative estimate of the harmonic pulse duration, as the radiating cusp moves forward with the velocity nearly equal to $c$, so that the expected harmonic pulse duration is much shorter.

**Estimation of the peak brightness**

The peak brightness of the experimental harmonic source [photons/(mm$^2$ mrad$^2$ s 0.1% bandwidth)] is estimated using the absolute photon number calculated as explained in the previous sections, the spectrograph acceptance angle of 2.4 mrad in the laser polarization plane and 12 mrad in the perpendicular direction, 13 fs harmonic pulse duration (conservative estimate), and the source area equals the area of ring, which diameter equals the self-focusing channel diameter (5 μm) and the thickness equals 1 μm (from the simulations). In the experiment, the polarization is linear, which means that only part of this ring emits harmonics, as seen in Fig. 3**a** of the main text. Also, we expect that the harmonic pulse is significantly shorter than the harmonic emission time, because the cusp moves with a velocity nearly equal to the driving laser pulse one. Thus, the peak brightness given in the main text is a conservative estimate, while the real value is to be determined in future experiments, where the source size and pulse duration are measured.

**Ruling out of previously known mechanisms of harmonic generation in gas jets**

**Atomic harmonics.**[44, 45] Due to the symmetry with respect to the electric field reversal, these harmonics are generated two times during each laser cycle, so the harmonic separation is $2\pi/(T_0/2) = 2\omega_0$, and only odd harmonics are generated. Addition of the second harmonic pulse breaks the symmetry, so even harmonics can also be generated,[46, 47] with the dependence on the intensity and delay of the second harmonic pulse. The second harmonic may accidentally be generated in the experiment, but its intensity and phase should depend on the parameters. However in all of the spectra taken with a broad range of parameters the harmonics show no variation in amplitude and shape between the even and odd orders. This means that the possible presence of a self-generated variable second harmonic does not take part in the high order harmonic generation process, which in turn means that atomic harmonics are not relevant to the experiment. There are additional reasons which allow excluding the atomic harmonics: (i) the laser irradiance is orders of magnitude larger than necessary for full He ionization; (ii) the base frequency is down-shifted, which happens only within the high-intensity region, where He is fully ionized; (iii) there is no strong sensitivity to backing pressure, which is important for the atomic harmonics phase-matching.

**Nonlinear Thomson scattering.** The nonlinear Thomson scattering[48-50] gives single-electron spectra with the calculated shapes[49, 50] which resemble the ones recorded in our experiment. Low-order[11, 12] and vacuum-ultraviolet[53] harmonics and continuum radiation[54] attributed to the nonlinear Thomson scattering have been observed experimentally. However, more detailed analysis shows that there are two spectral features in our experiment which cannot be explained by the nonlinear Thomson scattering. First, it is well known that along with the gradual downshift of the average laser frequency, the laser spectrum in plasma is broadened, which in the experiment is visible in the transmitted spectra (Fig. S5**a**), so that different parts of the pulse have different frequencies. Typically, the head of the pulse has a lower frequency than the tail. The nonlinear Thomson scattering in the field of such pulse indeed contains resolvable lower harmonic orders,[11, 53] but higher orders are blurred[54] due to the presence of different base frequencies. Note that the harmonic generation mechanism proposed in this paper is based on the radiation by an electron density cusp, which is a very localized charge, so that laser frequency does not change across the charge location and resolvable harmonics can be generated up to high orders, as it is the case in the experiment. Second, the number of photons recorded in our experiment cannot be explained by the nonlinear Thomson scattering. Using numerical calculations,[49] we obtained the spectra of radiation scattered by a single electron in the field of a laser pulse with an over-estimated amplitude $a_0 = 7$, at a non-zero observation angle $\theta = 15°$, which is close to the optimum emission angle.[49] Even for this amplitude and angle, at 100 eV the radiated energy is at most $\approx 2\times10^{-10}$ nJ/eV/sr. The number of electrons encountered by the self-focused laser pulse can be estimated as $N_e = n_e \, \Delta x \, \pi \, d_{sf}^2/4 \approx 2\times10^9$. Here we use the peak electron density $n_e = 2.7\times10^{19}$ cm$^{-3}$, the self-focusing channel diameter (S3) $d_{sf} = 5$ μm, and the harmonic emission length $\Delta x = 4$ μm. These give $\approx 0.4$ nJ/eV/sr at the photon energy of 100 eV, which is more than two orders of magnitude (~300 times) smaller than that observed in the experiment. Note that in this estimation, we over-estimate wherever possible the Thomson scattering signal and under-estimate the signal recorded in the experiment, because we do not take into account a possible optics contamination and absorption in the outer gas jet regions. More accurate estimation would give even a larger gap between nonlinear Thomson scattering and the experiment.

**Betatron radiation.**[55] In this case, the radiation consists of harmonics with the base frequency determined by the plasma frequency $\omega_{pe}$ and electron bunch gamma-factor $\gamma_e$. This cannot provide the $\omega_f$ values always close to the laser frequency, which is observed in the experiment.

## Particle-in-cell simulations

Three-dimensional particle-in-cell simulations were performed with the Relativistic Electro-Magnetic Particle-mesh (REMP) code[56] using SGI Altix 3700 supercomputer. In the simulations, the laser pulse propagates in a tenuous plasma along the *x*-axis. The pulse is linearly polarized in the direction of the *y*-axis; its shape is Gaussian and its full width at half-maximum is $10\lambda_0$ in every direction. The initial laser pulse dimensionless amplitude is $a = 6.6$. The plasma is fully ionized; the electron density is $n_e = 1.14 \times 10^{18} \text{cm}^{-3} \times (1 \mu\text{m}/\lambda_0)^2$. We consider interaction mainly near the location of the laser pulse, so that the response from ions can be neglected in comparison with much lighter electrons, i. e. the ion-to-electron mass ratio is assumed to be $m_i/m \rightarrow \infty$. The simulation grid dimensions are 4000×992×992 along the *x*, *y* and *z* axes; the grid mesh sizes are $dx = \lambda_0/32$, $dy = dz = \lambda_0/8$; total number of quasi-particles is $2.3 \times 10^{10}$.

**Estimation of the energy emitted by the electron density singularity (cusp)**

In the reference frame moving with the cusp, the total power emitted by $N_e$ electrons is[57, §73]

$$P'_{\text{cusp}} = \frac{e^2}{3c} N_e^2 a_0^2 \left(1 + \frac{3}{8} a_0^2\right) \omega_0^2 \approx \frac{e^2}{8c} N_e^2 a_0^4 \omega_0^2 \qquad (S5)$$

In the laboratory reference frame, this is multiplied by the gamma-factor corresponding to the group velocity of laser pulse, $\gamma$. Thus, for the total emitted energy we obtain

$$W \approx \frac{e^2}{8c} N_e^2 a_0^4 \gamma \omega_0^2 \tau . \qquad (S6)$$

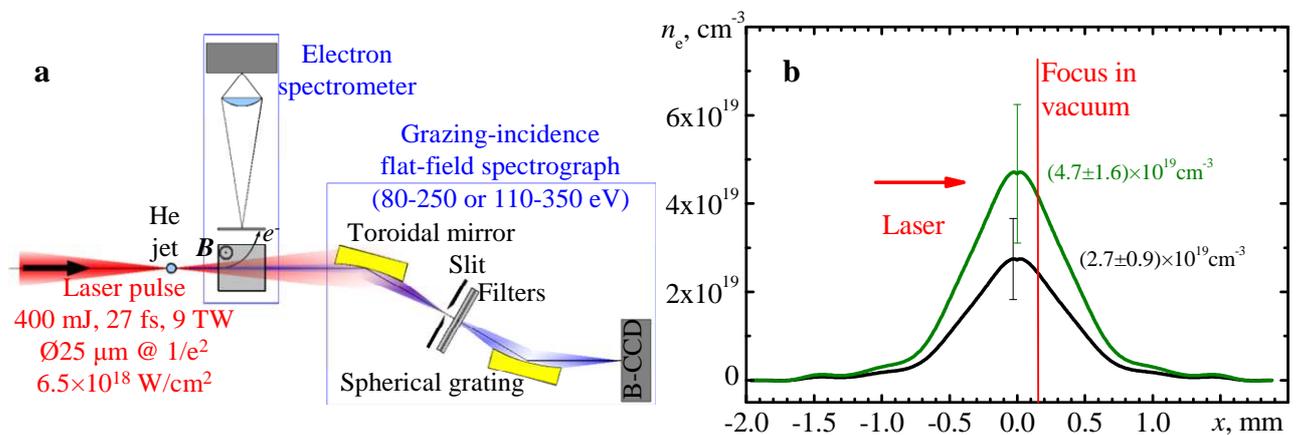

Fig. S1. **a** Experimental setup. A Ti:Sapphire laser pulse[58] (wavelength of 820 nm, energy of 0.4 J, duration of 27 fs Full Width at Half Maximum (FWHM) measured with a self-made Transient Grating FROG,[59] peak power of 9 TW, spot diameter of 25 μm at $1/e^2$, irradiance in vacuum of $6.5 \times 10^{18}$ W/cm², which corresponds to the amplitude $a_{0,vac}$ = 1.7)

irradiates a pulsed supersonic helium gas jet (conical nozzle with 1 mm diameter orifice, Mach number $M$ = 3.3). The harmonics in the soft X-ray region are measured along the laser propagation direction with a grazing-incidence flat-field spectrograph similar to the one described in[30, 60] comprising a gold-coated toroidal mirror with the main radii of 4897 mm and 23.74 mm operating at the incidence angle of 88°, a 200 μm slit, optical blocking filters, a spherical mechanically ruled flat-field grating[34] with the radius of 5649 mm and the nominal groove density of 1200 lines/mm operating at the incidence angle of 87°, and a 16 bit per pixel, 1100×330 back-illuminated Charge-Coupled Device (CCD) with the pixel size of 24 μm operated at -24°C. The acceptance angle of the spectrograph is $3\times10^{-5}$ sr. The estimated spatial resolution of the spectrograph is several tens of micrometers, limited by the geometrical aberrations of the grazing-incidence optics. Two spectral ranges have been used, 80-250 eV (employing two 160 nm Mo/C multilayer filters[33]) and 110-350 eV (with two 200 nm Pd filters). The electrons accelerated[61] in the gas jet are deflected by a permanent magnet (0.76 T, 10 cm × 10 cm) and directed to the phosphor screen (DRZ-High), which is imaged onto the gated intensified CCD to obtain the electron energy spectra in each shot.[62] In some shots there are additional two laser pulses used for the diagnostic and other purposes; we have checked (by blocking and time delays) that these additional pulses do not influence the processed described in this paper. **b** The electron density profiles at 1 mm above the nozzle (laser irradiation position) estimated from the neutral gas density assuming double ionization of He atoms; this assumption is justified by the high intensity of the laser pulse, which exceeds the threshold of the barrier suppression ionization by few orders of magnitude. The neutral gas density profiles are measured with the interferometry. The FWHM of the density distribution is 840 µm. The arrow and vertical line show the laser beam direction and the position of focus in vacuum, respectively. The error bars are due to the noise of CCD used for the interferometry, which affects the density reconstruction process.

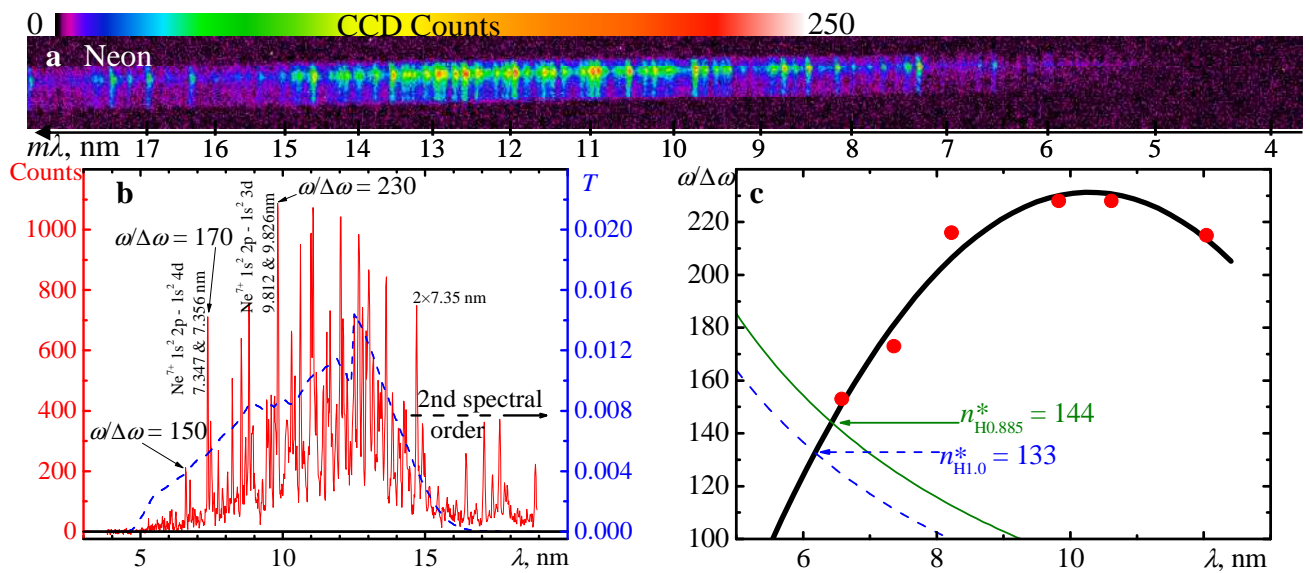

Fig. S2. **a** CCD raw data, spectrum of Ne ion emission used for the in-place spectrograph calibration; the Ne plasma is created by the same laser pulse in the same nozzle backed with Ne gas. **b** Solid line shows the lineout of the raw data shown in the frame **a** with the identification of some Ne ion lines and estimated resolution using the line widths. Dashed line shows the idealized spectrograph throughput, which is the product of the toroidal mirror reflectivity, filter transmission, grating efficiency, and CCD quantum efficiency. **c** The circles show the resolving power of the spectrograph $\omega/\Delta\omega$ estimated using the width of Ne ion lines. The thick solid line is the 2$^{\text{nd}}$ order polynomial fit to the data. The thin solid and dashed lines show the harmonic orders for the base frequencies of $\omega_0$ and $0.885\omega_0$. The crossings of these lines with the curve $\omega/\Delta\omega$ give the estimations of the maximum resolvable harmonic orders.

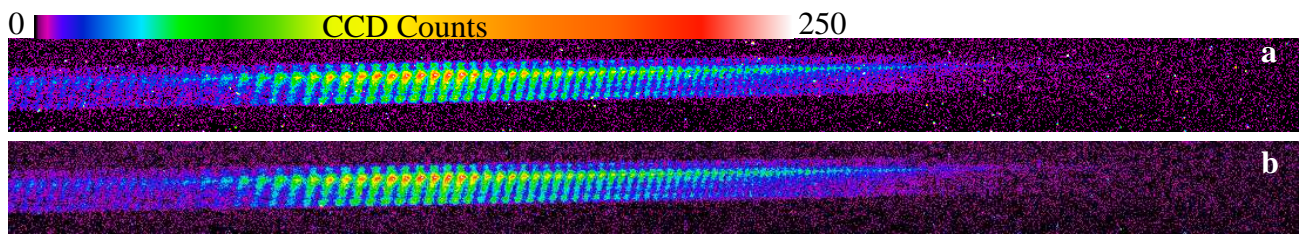

Fig. S3. Example of bright spot removal, the same data as in Fig. 1 of the main text. **a** Original data (no processing). **b** Image after the bright spot removal.

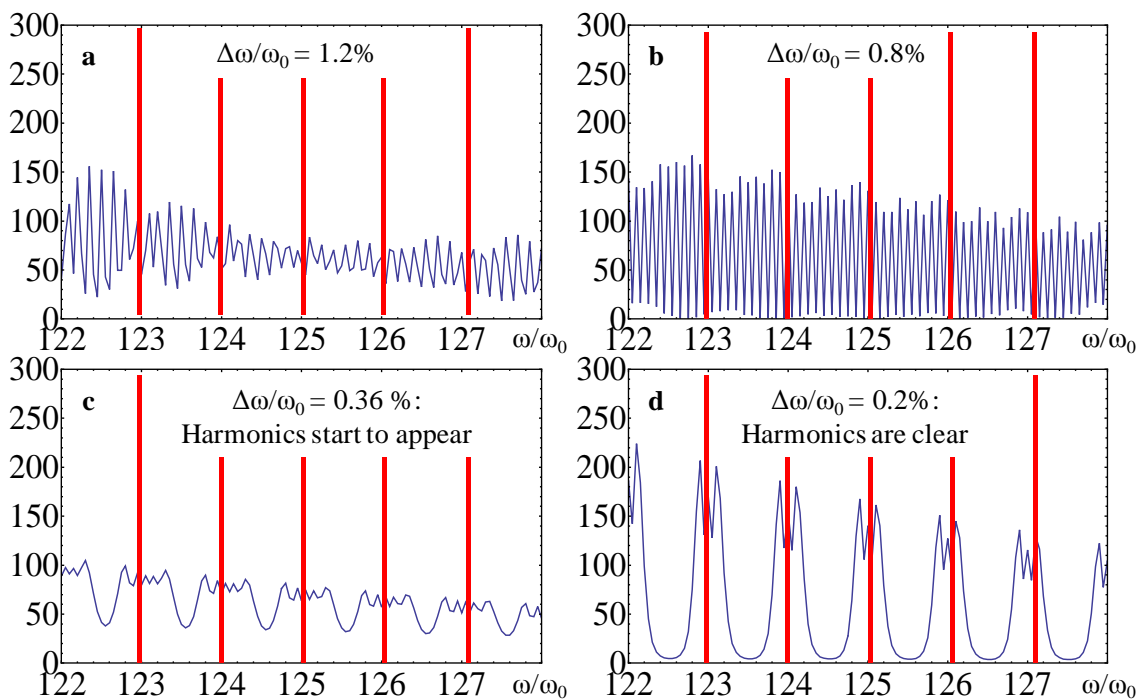

Fig. S4. A model example of harmonics generated by the laser pulse with gradually changing frequency. The electric field is calculated as
$$E = \sum_{n=110}^{140} 2^{-(n-110)/15} \cos[2\pi n\omega_0(1+\alpha t)t]$$ within the time range (-10, 10), so that the frequency varies from $(1-10\alpha)\omega_0$ to $(1+10\alpha)\omega_0$, $\Delta\omega/\omega_0 = 20\alpha$. The

harmonics around order $n_H{}^* = 126$ become distinguishable when the relative frequency change is smaller than $\approx 1/2n_H{}^* \approx 0.4\%$.

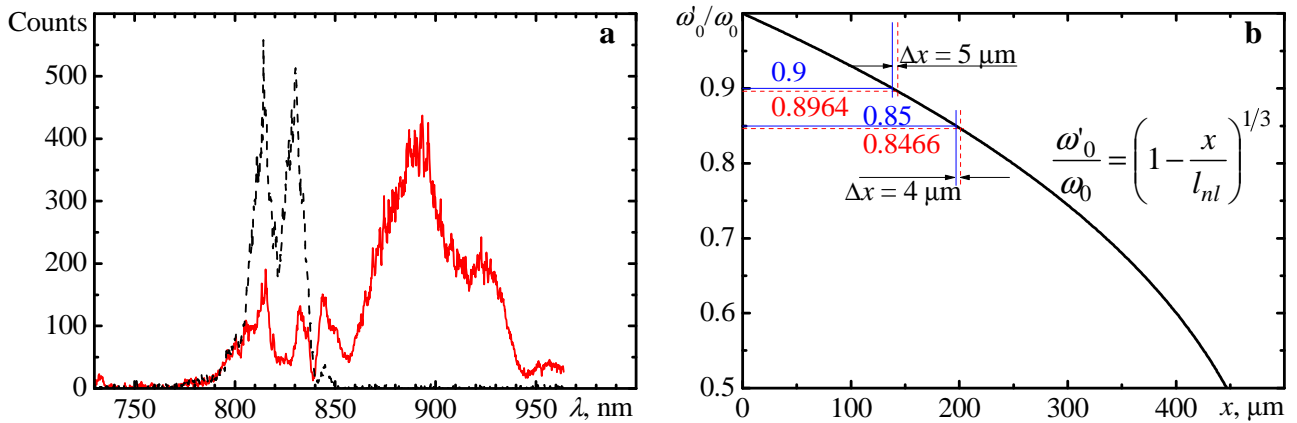

Fig. S5. Broadening of the transmitted laser spectrum and downshift of the average frequency due to nonlinear pulse depletion. **a** Solid line, the spectrum of transmitted radiation in the same shot as shown in Fig. 1 of the main text. Dashed line, the spectrum of laser obtained with the same setup, but without plasma. The radiation around 800 nm is suppressed by a dielectric-coated mirror so that the same setup can be used to record the spectrum of original laser pulse and significantly depleted and frequency downshifted pulse transmitted through plasma. **b** Average laser pulse frequency estimated using Eq. (S4) assuming uniform plasma density ($n_e = 2.7 \times 10^{19}$ cm$^{-3}$). The estimated values of the propagation distance $\Delta x$ during which the laser frequency changes by 0.4% are shown for two positions corresponding to the frequencies of $0.9\omega_0$ and $0.85\omega_0$, similar to those observed in the experiment. Note that the estimated $\Delta x$ value depends weakly on the position in plasma $x$ as long as $x < l_{nl}$ or, in other words, $\omega_0{}' \approx \omega_0$, which is the case in the experiment.